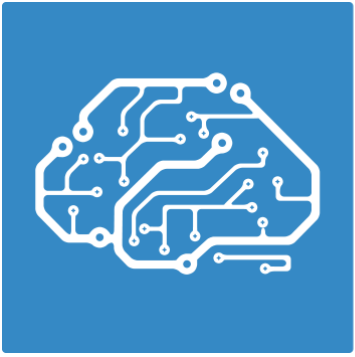

CBMM Memo No. 9                          April 26, 2014

# A role for recurrent processing in object completion: neurophysiological, psychophysical, and computational evidence

by

Hanlin Tang, Calin Buia, Joseph Madsen, William S. Anderson, Gabriel Kreiman

Abstract:

Recognition of objects from partial information presents a significant challenge for theories of vision because it requires spatial integration and extrapolation from prior knowledge. We combined neurophysiological recordings in human cortex with psychophysical measurements and computational modeling to investigate the mechanisms involved in object completion. We recorded intracranial field potentials from 1,699 electrodes in 18 epilepsy patients to measure the timing and selectivity of responses along human visual cortex to whole and partial objects. Responses along the ventral visual stream remained selective despite showing only 9-25% of the object. However, these visually selective signals emerged ~100 ms later for partial versus whole objects. The processing delays were particularly pronounced in higher visual areas within the ventral stream, suggesting the involvement of additional recurrent processing. In separate psychophysics experiments, disrupting this recurrent computation with a backward mask at ~75ms significantly impaired recognition of partial, but not whole, objects. Additionally, computational modeling shows that the performance of a purely bottom-up architecture is impaired by heavy occlusion and that this effect can be partially rescued via the incorporation of top-down connections. These results provide spatiotemporal constraints on theories of object recognition that involve recurrent processing to recognize objects from partial information.

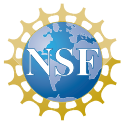

This work was supported by the Center for Brains, Minds and Machines (CBMM), funded by NSF STC award CCF-1231216.

**Introduction**

During natural viewing conditions, we often have access to only partial information about objects due to limited viewing angles, poor luminosity or object occlusion. Despite these difficulties, the visual system shows a remarkable ability to interpret objects from their constituent parts. How the visual system can recognize objects from limited information while still maintaining fine discriminability between like objects remains poorly understood and represents a significant challenge for computer vision algorithms and theories of vision.

Visual shape recognition is orchestrated by a cascade of signal processing steps along the ventral visual stream (for reviews, see (Connor et al., 2007; Logothetis and Sheinberg, 1996; Rolls, 1991; Tanaka, 1996)). Neurons in the highest echelons of the ventral stream in macaque monkeys, the inferior temporal cortex (ITC), demonstrate strong selectivity to complex objects (e.g. (Desimone et al., 1984; Hung et al., 2005; Ito et al., 1995; Keysers et al., 2001; Logothetis et al., 1995; Miyashita and Chang, 1988; Richmond et al., 1983; Rolls, 1991)). In the human brain, several areas within the occipital-temporal lobe showing selective responses to complex shapes have been identified using neuroimaging (Grill-Spector and Malach, 2004; Haxby et al., 1991; Kanwisher et al., 1997; Taylor et al., 2007) and invasive physiological recordings (Allison et al., 1999; Liu et al., 2009; Privman et al., 2007). Converging evidence from behavioral studies (Kirchner and Thorpe, 2006; Thorpe et al., 1996), human scalp electroencephalography (Thorpe et al., 1996), monkey (Hung et al., 2005; Keysers et al., 2001; Optican and Richmond, 1987) and human (Allison et al., 1999; Liu et al., 2009) neurophysiological recordings has established that selective responses to and rapid recognition of isolated whole objects can occur within 100 ms of stimulus onset. As a first-order approximation, the speed of visual processing suggests that initial recognition may occur in a largely feed-forward fashion, whereby neural activity progresses along the hierarchical architecture of the ventral visual stream with minimal contributions from feedback connections between areas or within-area recurrent computations (Deco and Rolls, 2004; Fukushima, 1980; LeCun et al., 1998; Riesenhuber and Poggio, 1999).

A critical feature of visual recognition is the remarkable degree of robustness to object transformations. Recordings in ITC of monkeys (Desimone et al., 1984; Hung et al., 2005; Ito et al., 1995; Logothetis and Sheinberg, 1996) and humans (Liu et al., 2009) have revealed a significant degree of tolerance to object transformations. Visual recognition of isolated objects under certain transformations such as scale or position changes do not incur additional processing time at the behavioral or physiological level (Allison et al., 1999; Biederman and Cooper, 1991; Desimone et al., 1984; Ito et al., 1995; Liu et al., 2009; Logothetis et al., 1995; Logothetis and Sheinberg, 1996) and can be described using purely bottom-up computational models. While bottom-up models may provide a reasonable approximation for rapid recognition of whole isolated objects, top-down as well as horizontal projections abound throughout visual cortex (Callaway, 2004; Felleman and Van Essen, 1991). The contribution of these projections to the strong robustness of object recognition to various transformations remains unclear. In particular, recognition of objects from partial information is a difficult problem for purely feed-forward architectures and may involve significant contributions from recurrent connections as shown in attractor networks (Hopfield, 1982; O'Reilly et al., 2013) or studies of Bayesian inference (Lee and Mumford, 2003).

Previous studies have examined the brain areas involved in pattern completion with human neuroimaging (Lerner et al., 2004; Schiltz and Rossion, 2006; Taylor et al., 2007), the degree of selectivity in physiological signals elicited by partial objects (Issa and Dicarlo, 2012; Kovacs et al., 1995b; Nielsen et al., 2006; Rutishauser et al., 2011) and delays associated with recognizing occluded or partial objects (Biederman, 1987; Brown and Koch, 2000; Johnson and Olshausen, 2005). Several studies have principally focused on *amodal completion*, i.e., the linking of disconnected parts to a single 'gestalt', using geometric shapes or line drawings and strong occluders that provided depth cues (Brown and Koch, 2000; Chen et al., 2010; Johnson and Olshausen, 2005; Murray et al., 2001; Nakayama et al., 1995; Sehatpour et al., 2008). Amodal completion is an important step in recognizing occluded objects. In addition to determining that different parts belong to a whole, the brain has to jointly process the parts to recognize the object (Gosselin and Schyns, 2001; Nielsen et al., 2006; Rutishauser et al., 2011), which we study here.

We investigated the spatiotemporal dynamics underlying object completion by recording field potentials from intracranial electrodes implanted in epilepsy patients while subjects recognized objects from partial information. Even with very few features present (9-25% of object area shown), neural responses in the ventral visual stream, principally in the inferior occipital gyrus and fusiform gyrus, retained object selectivity. Visually selective responses to partial objects emerged about 100ms later than responses to whole objects. These delays persisted when controlling for differences in contrast, signal amplitude, and the strength of selectivity. Furthermore, the processing delays associated with interpreting objects from partial information increased along the visual hierarchy. These delays stand in contrast to other object transformations such as position or scale changes that do not lead to physiological or behavioral delays. Together, these results argue against a feed-forward explanation for recognition of partial objects and provide evidence for the involvement of highest visual areas in recurrent computations orchestrating pattern completion.

**Results**

We recorded intracranial field potentials (IFPs) from 1,699 electrodes in 18 subjects (11 male, 17 right-handed, 8-40 years old) implanted with subdural electrodes to localize epileptic seizure foci. Subjects viewed images containing grayscale objects presented for 150 ms. After a 650 ms delay period, subjects reported the object category (animals, chairs, human faces, fruits, or vehicles) by pressing corresponding buttons on a gamepad (**Figure 1A**). In 30% of the trials, the objects were unaltered (referred to as the 'Whole' condition). In 70% of the trials, partial object features were presented through randomly distributed Gaussian "bubbles" (**Figure 1B, Experimental Procedures**, referred to as the 'Partial' condition) (Gosselin and Schyns, 2001). The number of bubbles was calibrated at the start of the experiment such that performance was ~80% correct. The number of bubbles (but not their location) was then kept constant throughout the rest of the experiment. For 12 subjects, the objects were presented on a gray background (the 'Main' experiment). While contrast was normalized across whole objects, whole objects and partial objects had different contrast levels because of the gray background. In 6 additional subjects, a modified experiment (the 'Variant' experiment) was performed where contrast was normalized between whole and partial objects by presenting objects on a background of phase-scrambled noise (**Figure 1B**).

The performance of all subjects was around the target correct rate (**Figure 1C**, 79%±7%, mean±SD). Performance was significantly above chance (Main experiment: chance = 20%, 5-alternative forced choice; Variant experiment: chance = 33%, 3-alternative forced choice) even when only 9-25% of the object was visible. As expected, performance for the whole condition was near ceiling (95±5%, mean±SD). The analyses presented throughout the manuscript were performed on correct trials only.

**Object selectivity was retained despite presenting partial information**

Consistent with previous studies, multiple electrodes showed strong visually selective responses to whole objects (Allison et al., 1999; Davidesco et al., 2013; Liu et al., 2009). An example electrode from the 'Main' experiment, located in the Fusiform Gyrus (see map of electrode locations in **Figure 4E**), had robust responses to several exemplars in the Whole condition, such as the one illustrated in the first panel of **Figure 2A**. These responses could also be observed in individual trials of face exemplars (gray traces in **Figure 2A-B,** left). This electrode was preferentially activated in response to faces compared to the other objects in the Whole condition (**Figure 2C**, left). Responses to stimuli other than human faces were also observed, such as the responses to several animal (red) and fruit (orange) exemplars (**Figure S1B**).

The neural responses in this example electrode were remarkably preserved in the Partial condition, where only 11±4% (mean±SD) of the object was visible. Despite the variability in bubble locations, robust responses were observed in single trials (**Figure 2A** and **Figure 2B,** right). Even when largely disjoint sets of features were presented, the evoked responses were similar (e.g., compare **Figure 2A**, third and fourth images). Because the bubble locations varied from trial to trial, there was significant variability in the latency of the visual response (**Figure 2B**, right); this variability affected the average responses to each category of partial objects (**Figure 2C**, right). Despite this variability, the electrode remained selective and kept the stimulus preferences at the category and exemplar level (**Figure 2C**).

The responses of an example electrode from the 'Variant' experiment support similar conclusions (**Figure 3**). Even though only 21%±4% (mean±SD) of the object was visible, the electrode demonstrated robust responses in single trials (**Figure 3A-B**), and strong selectivity both for whole objects and partial objects at the category level (**Figure 3C**). While the selectivity was consistent across single trials, there was significantly more trial-to-trial variation in the timing of the responses to partial objects compared to whole objects (**Figure 3B**, top right).

To measure the strength of selectivity, we employed two approaches. The first approach (denoted 'ANOVA') was a non-parametric one-way analysis of variance test to evaluate whether and when the variance in the IFP responses across categories was larger than the variance within a category. An electrode was denoted "selective" if, during 25 consecutive milliseconds, the ratio of variances across versus within categories (F-statistic) was greater than a significance threshold determined by a bootstrapping procedure to ensure a false discovery rate $q<0.001$ (F = 5.7) (**Figure 2-3D**). The ANOVA test evaluates whether the responses are statistically different when averaged across trials, but the brain needs to discriminate among objects in single trials. To evaluate the degree of selectivity in single trials, we employed a statistical learning approach to measure when information in the neural response became available to correctly classify the object into one of the five categories (denoted 'Decoding'; **Figure 2E**, chance = 20%; **Figure 3E**, chance = 33%). An electrode was considered "selective" if the decoding performance exceeded a threshold determined to ensure $q < 0.001$ (**Experimental Procedures**).

Of the 1,699 electrodes, 210 electrodes (12%) and 163 electrodes (10%) were selective during the Whole condition in the ANOVA and Decoding tests, respectively. To be conservative, we focused subsequent analyses only on those electrodes selective in both tests, yielding 113 electrodes (7% of the total number of electrodes), 83 from the main experiment and 30 from the variant (**Table 1**). As a control, shuffling the object labels yielded only 2.78±0.14 (mean±s.e.m., 1000 iterations) electrodes (0.16% of the total). Similar to previous reports, the preferred category of different electrodes spanned all five object categories, and the electrode locations were primarily distributed along the ventral visual stream (**Figure 4E-F**) (Liu et al., 2009). As demonstrated for the examples in **Figures 2** and **3**, even though only 9-25% of each object was shown, 30 electrodes (24%) remained visually selective in the Partial condition (Main experiment: 22; Variant experiment: 8) whereas the shuffling control yielded an average of 0.06 and 0.04 electrodes in the Main and Variant experiments respectively (**Table 1**).

The examples in **Figure 2C** and **3C** seem to suggest that the response amplitudes were larger in the Whole condition. However, this effect was due to averaging over trials and the increased trial-to-trial variability in the response latency for the Partial condition (i.e. no amplitude changes are apparent in the single trial data shown in **Figure 2B** and **3B**). For the 22 electrodes selective during both conditions in the Main experiment, the IFP amplitude of the responses in the preferred category, defined as the range of the IFP signal from 50 to 500 ms, was not significantly reduced (**Figure 4A**, p=0.68, Wilcoxon rank-sum test). The variability in the latency and the waveform, however, reduced the strength of category selectivity in the Partial condition, as measured by the F-statistic ($p<10^{-4}$, signed-rank test) and the decoding performance ($p<10^{-4}$, signed-rank test).

To compare different brain regions, we measured the percentage of electrodes in each gyrus that were selective in either the Whole condition or in both conditions (**Figure 4B-C,** see **Experimental Procedures** for electrode localization). Consistent with previous reports, electrodes in both early (Occipital Pole and Inferior Occipital Gyrus) and late (Fusiform Gyrus and Inferior Temporal Gyrus) visual areas were selective in the Whole condition (**Figure 4C**, black dots) (Allison et al., 1999; Davidesco et al., 2013; Liu et al., 2009). The locations with the highest percentages of electrodes selective to partial objects were primarily in higher visual areas, such as the Fusiform Gyrus and Inferior Occipital Gyrus (**Figure 4E**, gray bars, $p = 2\times10^{-6}$ and $5\times10^{-4}$ respectively, Fisher's exact test).

The observation that even non-overlapping sets of features can elicit robust responses (e.g., third and fourth panel in **Figure 2A**) suggests that the electrodes tolerated significant trial-to-trial variability in the visible object fragments. To quantify this observation across the population, we defined the percentage of overlap between two partial images of the same object by computing the number of pixels shared by the image pair divided by the object area (**Figure 4D**, insert). We considered partial images where the response to the preferred category was highly discriminable from the response to the non-preferred categories (Experimental Procedures). Even for these trials with robust responses, 45% of the image pairs (n = 10,438 total image pairs from the 22 electrodes in the Main experiment) had less then 5% overlap, and 11% of the pairs had less than 1% overlap (**Figure 4D**). Furthermore, in every electrode, there existed pairs of robust responses where the partial images had <1% overlap.

In sum, electrodes in the highest visual areas in the human ventral stream retained visual selectivity to partial objects, their responses could be driven by disjoint sets of object parts and the response amplitude but not the degree of selectivity was similar to that of whole objects.

**Delayed responses to partial objects**

In addition to the changes in selectivity described above, the responses to partial objects were delayed compared to the corresponding responses to whole objects (e.g. compare Whole versus Partial in the single trial responses in **Figure 2A-B** and **3A-B**). To compare the latencies of responses to Whole and Partial objects, we measured both selectivity latency and visual response latency. Selectivity latency indicates when sufficient information becomes available to distinguish among different objects or object categories, whereas the response latency denotes when the visual response differs from baseline (**Experimental Procedures**).

Quantitative estimates of latency are difficult because they depend on multiple variables, including number of trials, response amplitudes and thresholds. Here we independently applied different measures of latency to the same dataset. The selectivity latency in the responses to whole objects for the electrode shown in **Figure 2** was 100±8 ms (mean ± 99% CI) based on the first time point when the F-statistic crossed the statistical significance threshold (**Figure 2D**, black arrow). The selectivity latency for the partial objects was 320±6 ms (mean ± 99% CI), a delay of 220 ms. A comparable delay of 180 ms between partial and whole conditions was obtained using the single-trial decoding analyses (**Figure 2E**). Similar delays were apparent for the example electrode in **Figure 3**.

We considered all electrodes in the Main experiment that showed selective responses to both whole objects and partial objects (n=22). For the responses to whole objects, the median latency across these electrodes was 155 ms, which is consistent with previous estimates (Allison et al., 1999; Liu et al., 2009). The responses to partial objects showed a significant delay in the selectivity latency as measured using ANOVA (median latency difference between Partial and Whole conditions = 117 ms, **Figure 4E,** black dots, $p < 10^{-5}$) or Decoding (median difference = 158 ms, **Figure 4F**, black dots, $p < 10^{-5}$).

We examined several potential factors that might correlate with the observed latency differences. Stimulus contrast is known to cause significant changes in response magnitude and latency across the visual system (e.g. (Reich et al., 2001; Shapley and Victor, 1978)). As noted above, there was no significant difference in the response magnitudes between Whole and Partial conditions (**Figure 4A**). To further investigate whether contrast could explain the physiological delays observed in the Partial condition, we examined the experimental variant where the images had the same contrast in both Whole and Partial conditions (**Figure 1B**). In this Variant experiment, we still observed latency differences between conditions (median difference = 73 ms (ANOVA), **Figure 4E,** and median difference = 93 ms (Decoding), **Figure 4F**, gray circles).

We asked whether the observed delays could be related to differences in the IFP response strength or the degree of selectivity by conducting an analysis of covariance (ANCOVA). ANCOVA is a general linear model that tests for the significance of an effect while controlling for the variance contributed by other factors. The latency difference between conditions, as measured with the F-statistic, was significant even when accounting for differences in IFP amplitude ($p < 10^{-9}$) or strength of selectivity ($p < 10^{-8}$). Even though the average amplitudes were similar for whole and partial objects (**Figure 4A**), the variety of partial images could include a wider distribution with weak stimuli that failed to elicit a response. To evaluate whether such potential weaker responses could contribute to the latency differences, we identified those trials where the decoder was correct at 500ms and evaluated the decoding dynamics before 500 ms under these matched performance conditions. The latency difference between whole and partial conditions was still statistically significant when matching decoding performance between conditions ($p<10^{-7}$).

Differences in eye movements between whole and partial conditions could potentially contribute to latency delays. We minimized the impact of eye movements by using a small stimulus size (5 degrees), fast presentation (150 ms) and trial order randomization. Furthermore, we recorded eye movements along with the neural responses in two subjects. There were no clear differences in eye movements between whole versus partial objects in these two subjects, and those subjects contributed 5 of the 22 selective electrodes in the Main experiment. We also recorded eye movements from 20 healthy volunteers and found no difference in the statistics of saccades and fixation between Whole and Partial conditions.

Several studies have documented visual selectivity in different frequency bands of the IFP responses including broadband and gamma band signals (Davidesco et al., 2013; Liu et al., 2009; Vidal et al., 2010). We computed the power in the Gamma band (70-100 Hz) using a Hilbert Transform, and applied the previously described ANOVA and Decoding analyses. The response delays during the Partial condition documented above for the broadband signals were also observed when measuring the selectivity latency in the 70-100 Hz frequency band. The median latency difference between the Partial and Whole conditions in the Gamma band was 157 ms (70-100 Hz, n = 14).

Because the spatial distribution of bubbles varied from trial to trial, each image in the Partial condition revealed different visual features. To account for the stimulus heterogeneity, we also measured the latency of the visual response in each individual trial by determining when the IFP amplitude exceeded a threshold set as three standard deviations above the baseline activity. The latency differences between Whole and Partial conditions were apparent even in single trials (e.g. **Figure 2A, 2F**). These latency differences depended on the sets of features revealed on each trial. When we presented repetitions of partial objects with one fixed position of bubbles (the 'Partial Fixed' condition). Under those conditions, the IFP timing was more consistent across trials (**Figure 3C**, right bottom), but the latencies were still longer for partial objects than for whole objects.

The average response latencies in the Whole and Partial condition for the preferred category for the first example electrode were 172 and 264 ms respectively; the distributions in the two groups differed significantly (**Figure 2F**, Wilcoxon rank-sum test, $p < 10^{-6}$). The distribution of response latencies in the Whole condition was highly peaked (**Figure 2F, 3F**), whereas the distribution of latencies in the Partial condition showed a larger variation, driven by the distinct visual features revealed in each trial.

Across the population, delays were observed in the visual response latencies (**Figure 5A**, rank-sum test, $p < 10^{-15}$).

Delays in the response latency between Partial and Whole conditions had a distinct spatial distribution: most of the delays occurred in higher visual areas such as the fusiform gyrus and inferior temporal gyrus (**Figure 5B**). The latency difference was smaller for electrodes in early visual areas (occipital cortex) versus late visual areas (temporal lobe), as shown in **Figure 5C** ($p=0.02$, t-test). There was also a significant correlation between the latency difference and the electrode position along the anterior-posterior axis of the temporal lobe (Spearman's correlation = 0.43, permutation test, $p = 0.02$).

The analyses presented thus far only measured selectivity latency at the level of individual electrodes, but the subject has access to activity across many regions. To assess selectivity at a population level, we combined information from multiple electrodes and across subjects by constructing pseudopopulations (Hung et al., 2005) (**Experimental Procedures**). Decoding performance using electrode ensembles was both fast and accurate (**Figure 6C**). Category information emerged within 150 ms for whole objects (black thick line) and 260 ms for partial objects (gray thick line), and reached 80% and 45% correct rate, respectively (chance = 20%). Even for the more difficult problem of identifying the stimulus exemplar (chance = 4%), decoding performance emerged within 135 ms for whole objects (black dotted line) and 273 ms for partial objects (gray dotted line). Exemplar decoding accuracy reached 61% for whole objects and 14% for partial objects. Together, these results suggest that, within the sampling limits of our techniques, electrode ensembles also show delayed selectivity for partial objects.

In sum, we have independently applied several different estimates of latency that use statistical (ANOVA), machine learning (Decoding), or threshold (Response latency) techniques. These latency measures were estimated while taking into account changes in contrast, signal strength and degree of selectivity. Each definition of latency requires different assumptions and emphasizes different aspects of the response, leading to variations in the absolute values of the latency estimates. Yet, independently of the specific definition, the latencies for partial objects were consistently delayed with respect to the latencies to whole objects (the multiple analyses are summarized in **Figure 6A**).

In contrast to the processing of whole objects, which can be explained by a feed-forward architecture, the increased latencies observed with partial objects suggests a role for recurrent or top-down computations. We hypothesized that disrupting such recurrent computations would affect behavior for partial, but not whole, objects. To this end, we performed a separate backward masking experiment on healthy subjects. Backward masking is thought to disrupt recurrent processing in the ventral visual stream (**Discussion**). This psychophysics experiment was similar to the physiology experiment except for two main differences. First, the images were presented for variable times, ranging from 33 to 150 ms, which we refer to as the stimulus-onset asynchrony (SOA). Second, the images were followed by either a gray screen (unmasked condition) as before, or a phase-scrambled mask (masked condition) (**Figure 7A**). For whole objects, the backward mask did not significantly affect performance (**Figure 7B**, black solid line versus black dotted line). For occluded objects, however, performance decreased significantly with a backward mask for short SOAs compared to the unmasked condition (**Figure 7B,** gray solid line versus gray dotted line). For SOAs longer than 100 ms, the mask lost its efficacy.

Finally, we evaluated the performance of a purely bottom-up architecture in recognizing the same set of objects and partial objects. We considered the HMAX architecture as implemented in (Serre et al., 2007). We used the same 25 images presented in the psychophysics experiment (without masking) and in the physiology experiment and used an SVM classifier to decode the identity of the image (Hung et al., 2005; Serre et al., 2007). The results of these analyses are presented in **Figure 8B,** red curve. While the purely bottom-up architecture could well identify the objects under reduced amounts of occlusion, its performance dropped significantly with increasing occlusion. Impairment was notable even under amounts of occlusion that would not lead to decreased psychophysical performance. As a proof-of-principle demonstration, we implemented an attractor network (schematically illustrated in **Figure 8A**) at the top of the bottom-up architecture. The addition of these recurrent connections to the bottom-up architecture increased performance in recognition of occluded images (**Figure 8B**, blue curve).

**Discussion**

The visual system must maintain selectivity to individual objects while remaining tolerant to a myriad of transformations of those objects. Our results show that neural activity in the human occipitotemporal cortex remained visually selective (e.g. **Figure 2**) even when limited partial information about each object was presented (on average, only 18% of each object was visible). Despite the trial-to-trial variation in the features presented, the field potential response waveform, amplitude and object preferences were similar between the Whole and Partial conditions (**Figures 2-4**). The neural responses to partial objects required approximately 100 ms of additional processing time compared to whole objects (**Figures 4-6**). While the exact value of this delay may depend on stimulus parameters and task conditions, the requirement for additional computation was robust to a variety of different definitions of latencies including single-trial analyses, different frequency bands and different statistical comparisons (**Figure 6**) and persisted when accounting for changes in image contrast, signal strength, and the strength of selectivity. This additional processing time was more pronounced in higher areas of the temporal lobe including inferior temporal cortex and the fusiform gyrus than in earlier visual areas (**Figure 5B**).

Studies of object completion typically fall into two groups in terms of the stimuli used. Neurophysiological recordings in macaque IT have described neurons whose selectivity to simple geometric shapes is relatively invariant to occlusion (Kovacs et al., 1995b). Yet, other studies using more naturalistic stimuli with textures show that IT neuronal activity depends on the diagnostic value of the occluded parts (Issa and Dicarlo, 2012; Nielsen et al., 2006). These different findings illustrate the potential differences between amodal completion or line closure processes, and object completion arising from integrating information from partial textures. Our stimuli and findings address this second group, and do not test amodal completion. Both types of information are important to recognizing occluded objects, and whether they are orchestrated by similar or perhaps different mechanisms remains an important question for future work.

The observed speed of initial selective responses to presentation of whole objects is consistent with a largely bottom-up cascade of processes leading to recognition (Deco and Rolls, 2004; Fukushima, 1980; LeCun et al., 1998; Riesenhuber and Poggio, 1999; Rolls, 1991)(Hung et al., 2005; Keysers et al., 2001; Liu et al., 2009; Optican and Richmond, 1987; Thorpe et al., 1996). For partial objects, however, visually selective responses were significantly delayed with respect to whole objects (**Figures 5-6**). These physiological delays argue against a purely bottom-up signal cascade, and stand in contrast to other transformations (scale, position, rotation) that do not induce additional neurophysiological delays (Desimone et al., 1984; Ito et al., 1995; Liu et al., 2009; Logothetis et al., 1995; Logothetis and Sheinberg, 1996).

Delays in response timing have been used as an indicator for recurrent computations and/or top-down modulation (Buschman and Miller, 2007; Keysers et al., 2001; Lamme and Roelfsema, 2000; Schmolesky et al., 1998). In line with these arguments, we propose that the additional processing time implied by the delayed physiological responses can be ascribed to recurrent computations that rely on prior knowledge about the objects to be recognized (Ahissar and Hochstein, 2004). Anatomical studies have demonstrated extensive horizontal and top-down projections throughout visual cortex that could instantiate such recurrent computations (Callaway, 2004; Felleman and Van Essen, 1991). Several areas where such top-down and horizontal connections are prevalent showed selective responses to partial objects in our study (**Figure 4B-C**).

It is unlikely that these delays were due to the selective signals to partial objects propagating at a slower speed through the visual hierarchy in a purely feed-forward fashion. Selective electrodes in earlier visual areas did not have a significant delay in the response latency, which argues against latency differences being governed purely by low-level phenomena. Delays in the response latency were larger in higher visual areas, suggesting that top-down and/or horizontal signals within those areas of the temporal lobe are important for pattern completion (**Figure 6B**). Additionally, feedback is known to influence responses in visual areas within 100-200ms after stimulus onset, as evidenced in studies of attentional modulation that involve top-down projections (Davidesco et al., 2013; Lamme and Roelfsema, 2000; Reynolds and Chelazzi, 2004). Those studies report onset latencies of feedback similar to the delays observed here in the same visual areas along the ventral stream.

The selective responses to partial objects were not exclusively driven by a single object patch (**Figure 2A-B, 3A-B**). Rather, they were tolerant to a broad set of partial feature combinations. While our analysis does not explicitly rule out common features shared by different images with largely non-overlapping pixels, the large fraction of trials with images with low overlap that elicited robust and selective responses makes this explanation unlikely (**Figure 4D**). The response latencies to partial objects were dependent on the features revealed: when we fixed the location of the bubbles, the response timing was consistent from trial to trial (**Figure 3B**).

A role for recurrent computations is supported by the psychophysics experiments involving backward masking (**Figure 7**). Neurophysiological studies in V1 (Macknik and Livingstone, 1998), ITC (Kovacs et al., 1995a; Rolls et al., 1999), and frontal eye fields (Thompson and Schall, 1999) suggest that the onset of the backward mask may disrupt any residual information about the preceding image in early visual areas, causing a mismatch between bottom up inputs and top-down signals from higher visual areas. This interpretation is consistent with the observation that backward masking caused more impairment in recognition of occluded objects compared to whole objects, particularly at short stimulus onset asynchrony values (**Figure 7**). These results suggest that recognizing occluded objects could be implemented via recurrent processing, and that disrupting this processing directly affects behavior. Backward masking is also known to disrupt the feed-forward processing of objects, but such effects occur at much smaller SOAs than the ones considered here (Felsten and Wasserman, 1980).

The distinction between purely bottom-up processing and recurrent computations confirms predictions from computational models of visual recognition and attractor networks. Whereas recognition of whole objects has been successfully modeled by purely bottom-up architectures (Deco and Rolls, 2004; Fukushima, 1980; LeCun et al., 1998; Riesenhuber and Poggio, 1999), those models struggle to identify objects with only partial information (Johnson and Olshausen, 2005; O'Reilly et al., 2013). Instead, computational models that are successful at pattern completion involve recurrent connections (Hopfield, 1982; Lee and Mumford, 2003; O'Reilly et al., 2013). Different computational models of visual recognition that incorporate recurrent computations include connections within the ventral stream (e.g. from ITC to V4) and/or from pre-frontal areas to the ventral stream. Our results implicate higher visual areas (**Figure 4C, 5B**) as participants in the recurrent processing network involved in recognizing objects from partial information. Additionally, the object-dependent and unimodal distribution of response latencies to partial objects (e.g. **Figure 2F**) suggest models that involve graded evidence accumulation as opposed to a binary switch.

Recognizing objects from partial information involves extrapolation and evaluating the extent to which the fragments are consistent with a stored representation of the whole object. Attractor networks have been shown to be able to solve the problem of pattern completion. In a typical implementation, neurons in the network communicate in all-to-all fashion with a symmetrical connectivity matrix (Hopfield, 1982). Under these conditions, it is possible to define the weights so that the dynamics of the network is described by an energy function, which decreases monotonically and is bounded below, converging onto attractor states. Starting the network at states that represent partial information leads to convergence to the attractors and hence pattern completion. A simple demonstration of how this could work in the context of visual recognition is presented in Figure 8. The additional recurrent computations implied by this network are consistent with the physiological and behavioral delays demonstrated here.

The current observations highlight the need for dynamical models of recognition to describe where, when and how recurrent processing interacts with feed-forward signals to describe object completion. Our findings provide spatial and temporal bounds to constrain these models. Such models should achieve recognition of objects from partial information within 200 to 300 ms, demonstrate delays in the visual response that are feature-dependent, and include a graded involvement of recurrent processing in higher visual areas. We speculate that the proposed recurrent mechanisms may be employed not only in the context of object fragments but also in visual recognition for other types of transformations that impoverish the image or increase task difficulty. The behavioral and physiological observations presented here suggest that the involvement of recurrent computations during object completion, involving horizontal and top-down connections, result in a representation of visual information in the highest echelons of the ventral visual stream that is selective and robust to a broad range of possible transformations.

**Acknowledgements**

We thank all the patients for their cooperation in participating in this study. We also thank Laura Groomes for assistance with the electrode localization and psychophysics experiments, and Sheryl Manganaro, Jack Connolly, Paul Dionne and Karen Walters for technical assistance. We thank Ken Nakayama and Dean Wyatte for comments on the manuscript.

**Author Contributions**

H.T. and C.B. performed the experiments and analyzed the data; H.T., C.B. and G.K. designed the experiments; R.M., N.E.C., W.S.A., and J.R.M. assisted in performing the experiments; W.S.A. and J.R.M. performed the neurosurgical procedures; H.T. and G.K. wrote the manuscript. All authors commented and approved the manuscript.

# Experimental Procedures

**Physiology Subjects**

Subjects were 18 patients (11 male, 17 right-handed, 8-40 years old, **Table S1**) with pharmacologically intractable epilepsy treated at Children's Hospital Boston (CHB), Brigham and Women's Hospital (BWH), or Johns Hopkins Medical Institution (JHMI). They were implanted with intracranial electrodes to localize seizure foci for potential surgical resection. All studies described here were approved by each hospital's institutional review boards and were carried out with the subjects' informed consent. Electrode locations were driven by clinical considerations; the majority of the electrodes were not in the visual cortex.

**Recordings**

Subjects were implanted with 2mm diameter intracranial subdural electrodes (Ad-Tech, Racine, WI, USA) that were arranged into grids or strips with 1 cm separation. Each subject had between 44 and 144 electrodes (94±25, mean±SD), for a total of 1,699 electrodes. The signal from each electrode was amplified and filtered between 0.1 and 100 Hz with sampling rates ranging from 256 Hz to 1000 Hz at CHB (XLTEK, Oakville, ON, Canada), BWH (Bio-Logic, Knoxville, TN, USA) and JHMI (Natus, San Carlos, CA and Nihon Kohden, Tokyo, Japan). A notch filter was applied at 60 Hz. All the data were collected during periods without any seizure events. In two subjects, eye positions were recorded simultaneously with the physiological recordings (ISCAN, Woburn, MA).

**Neurophysiology experiments**

After 500 ms of fixation, subjects were presented with an image (256x256 pixels) of an object for 150 ms, followed by a 650 ms gray screen, and then a choice screen (**Figure 1A**). The images subtended 5 degrees of visual angle. Subjects performed a 5-alternative forced choice task, categorizing the images into one of five categories (animals, chairs, human faces, fruits, or vehicles) by pressing corresponding buttons on a gamepad (Logitech, Morges, Switzerland). No correct/incorrect feedback was provided. Stimuli consisted of contrast-normalized grayscale images of 25 objects, 5 objects in each of the aforementioned 5 categories. In approximately 30% of the trials, the images were presented unaltered (the 'Whole' condition). In 70% of the trials, the visual features were presented through Gaussian bubbles of standard deviation 14 pixels (the 'Partial' condition, see example in **Figure 1B**) (Gosselin and Schyns, 2001). The more bubbles, the more visibility. Each subject was first presented with 40 trials of whole objects, then 80 calibration trials of partial objects, where the number of bubbles was titrated in a staircase procedure to set the task difficulty at ~80% correct rate. The number of bubbles was then kept constant throughout the rest of the experiment. The average percentage of the object shown for each subject is reported in **Figure 1C**. Unless otherwise noted (below), the positions of the bubbles were randomly chosen in each trial. The trial order was pseudo-randomized.

The contrast of the objects was normalized across the 25 exemplars in the Whole condition. However, due to the random positioning of the bubbles, the contrast could change across different trials in the Partial condition. To evaluate the extent to which these differences could contribute to the results, 6 of the 18 subjects performed a variant of the main experiment with three key differences. First, contrast was normalized between the Whole and Partial conditions by presenting all objects in a phase-scrambled background (**Figure 1B**). Second, in 25% of the Partial condition trials, the spatial distribution of the bubbles was fixed to a single seed (the 'Partial Fixed' condition). Each of the images in these trials was identical across repetitions. Third, because experimental time was limited, only objects from three categories (animals, human faces and vehicles) were presented to collect enough trials in each condition.

**Psychophysics experiments**

The experiment used while collecting neurophysiological data was not designed to collect behavioral reaction time data. We conducted a separate psychophysics test on 10 healthy volunteers (6 male, 10 right-handed) (**Figure 7**). The same stimulus set of 25 objects used in the neurophysiology experiments was shown in either the Whole or Occluded condition, and subjects categorized the object by pressing corresponding buttons on a gamepad. Eye location was recorded using an infrared camera eye tracker (EyeLink, SR Research, Mississauga, Canada). Each trial was initiated by fixating on a cross. Each image was presented for a variable time (33 ms, 50 ms, 100 ms or 150 ms), which we denote as the stimulus onset asynchrony (SOA). For half the trials, the image presentation was followed by a gray screen for 500 ms, and then a choice screen. For the other half, the image presentation was followed by a phase-scrambled mask for 500 ms, and then a choice screen. The experiment consisted of 1,200 trials, and lasted approximately one hour.

**Psychophysics eye-tracking**

During the physiological recordings, we collected eye tracking data for two subjects. To further evaluate the type of eye movements that subjects execute under the same experimental conditions, we conducted a separate psychophysics test on 20 healthy volunteers (8 male, 15 right-handed). These subjects completed the same two experiments (10 subjects, Main Experiment, 10 subjects, Variant experiment). Eye location was recorded using an infrared camera eye tracker (EyeLink, SR Research, Mississauga, Canada). The experiment consisted of 1,200 trials, and lasted approximately one hour. We did not record physiological data from these additional subjects.

**Data Analyses**

*Electrode Localization*

Electrodes were localized by co-registering the preoperative magnetic resonance imaging (MRI) with the postoperative computer tomography (CT) (Destrieux et al., 2010; Liu et al., 2009). For each subject, the brain surface was reconstructed from the MRI and then assigned to one of 75 regions by Freesurfer. Each surface was also co-registered to a common brain for group analysis of electrode locations. In **Figure 5B**, we computed the Spearman's correlation coefficient between the latency differences (Partial - Whole) and distance along the posterior-anterior axis of the temporal lobe. In **Figure 4C**, we partitioned the electrodes into three groups: Fusiform Gyrus, Inferior Occipital Gyrus, and Other. We used the Fisher's exact test to assess whether the proportion of electrodes selective in both conditions is greater in the Fusiform Gyrus versus Other, and in Inferior Occipital Gyrus versus Other.

*Visual response selectivity*

All analyses in this manuscript used correct trials only. Noise artifacts were removed by omitting trials where the intracranial field potential (IFP) amplitude exceeded five times the standard deviation. The responses from 50 to 500 ms after stimulus onset were used in the analyses.

*ANOVA*. We performed a non-parametric one-way analysis of variance (ANOVA) of the IFP responses. For each time bin, the F-statistic (ratio of variance across object categories to variance within object categories) was computed on the IFP response (Keeping, 1995). Electrodes were denoted 'selective' in this test if the F-statistic crossed a threshold (described below) for 25 consecutive milliseconds (e.g. **Figure 2D**). The latency was defined as the first time of this threshold crossing. The number of trials in the two conditions (Whole and Partial) was equalized by random subsampling; 100 subsamples were used to compute the average F-statistic. A value of 1 in the F-statistic indicates no selectivity (variance across categories comparable to variance within categories) whereas values above 1 indicate increased selectivity.

**Decoding**. We used a machine learning approach to determine if, and when, sufficient information became available to decode visual information from the IFP responses in single trials (Bishop, 1995). For each time point $t$, features were extracted from each electrode using Principal Component Analysis (PCA) on the IFP response from [50 $t$] ms, and keeping those components that explained 95% of the variance. The features set also included the IFP range (max – min), time to maximum IFP, and time to minimum IFP. A multi-class linear discriminant classifier with diagonal covariance matrix was used to either categorize or identify the objects. Ten-fold stratified cross-validation was used to separate the training sets from the test sets to avoid overfitting. The proportion of trials where the classifier was correct in the test set is denoted the 'Decoding Performance' throughout the text (e.g. **Figure 2E**). In the Main experiment, a decoding performance of 20% (1/5) indicates chance for categorization and 4% (1/25) indicates chance for identification. The dataset sizes in the Whole and Partial conditions were equalized by subsampling; we computed the average Decoding Performance across 100 different subsamples. An electrode was denoted 'selective' if the decoding performance crossed a threshold (described below) at any time point *t*, and the latency was defined as the first time of this threshold-crossing.

**Pseudopopulation**. Decoding performance was also computed from an ensemble of electrodes across subjects by constructing a pseudopopulation, and then performing the same analyses described above (**Figure 6A**). The pseudopopulation pooled responses across subjects (Hung et al., 2005; Mehring et al., 2003; Pasupathy and Connor, 2002). It should be noted that such pooling involves several assumptions including similarities across subjects and ignoring trial-to-trial correlations across electrodes (for discussion, see (Meyers and Kreiman, 2011)). The features for each trial in this pseudopopulation were generated by first randomly sampling exemplar-matched trials without replacement for each member of the ensemble, and then concatenating the corresponding features. The pseudopopulation size was set by the minimum dataset size of the subject, which in our data was 100 trials (4 from each exemplar). Because of the reduced data set size, four-fold cross-validation was used.

**Significance Thresholds.** The significance thresholds for ANOVA, Decoding and d', were determined by randomly shuffling the category labels 10,000 times, and using the value of the 99.9 percentile (ANOVA: $F = 5.7$, Decoding: 23%, d' = 0.7). This represents a false discovery rate $q = 0.001$. As discussed in the text, we further restricted the set of electrodes by considering the conjunction of the ANOVA and Decoding tests. We evaluated this threshold by performing an additional 1,000 shuffles and measuring the number of selective electrodes that passed the same selectivity criteria by chance. In **Table 1**, we present the number of electrodes that passed each significance test and the number of electrodes that passed the same tests after randomly shuffling the object labels.

*Latency Measures*

We considered several different metrics to quantify the selectivity latency (i.e. the first time point when selective responses could be distinguished), and the visual response latency (i.e. the time point when a visual response occurred). These measures are summarized in **Figure 6**.

**Selectivity latency.** The selectivity latency represented the first time point when different stimuli could be discriminated and was defined above for the ANOVA and Decoding analyses.

**Response Latency**

Latency of the visual response was computed at a per-trial level by determining the time, in each trial, when the IFP amplitude exceeded 3 standard deviations above the baseline activity. Only trials corresponding to the preferred category were used in the analysis. To test the multimodality of the distribution of response latencies, we used Hartigan's dip test. In 27 of the 30 electrodes, the unimodality null hypothesis could not be rejected ($p > 0.05$).

*Frequency Band Analyses*

Power in the Gamma frequency band (70-100 Hz) was evaluated by applying a 5th order Butterworth filter bandpass, and computing the magnitude of the analytical representation of the signal obtained with the Hilbert transform. The same analyses (ANOVA, Decoding, Per-Trial Latency) were applied to the responses from all electrodes in different frequency bands.

*Bubble Overlap Analyses*

For each pair of partial object trials, the percent of overlap was computed by dividing the number of pixels that were revealed in both trials by the area of the object (**Figure 4D**). Because low degree of object overlap would be expected in trials with weak physiological responses, we focused on the most robust responses for these analyses by considering those trials when the IFP amplitude was greater than the 90th percentile of the distribution of IFP amplitudes of all the non-preferred category trials. Note that this analysis includes all 22 electrodes in the main experiment, even though 12 of the 22 electrodes had either too few trials or too many bubbles to generate enough low overlap pairs for this analysis. Exclusion of those electrodes would further reinforce the conclusions in the text by increasing the percentage of discriminable pairs with <5% and <1% overlap to 73% and 27%, respectively.

**Table 1**

| Experiment | Frequency Band | Whole | Shuffled | Both | Shuffled |
|---|---|---|---|---|---|
| Main | Broadband | 83 | (1.66±0.07) | 22 | (0.06±0.01) |
| Variant | Broadband | 30 | (1.12±0.12) | 8 | (0.04±0.03) |
| Main | Gamma | 53 | (1.56±0.05) | 14 | (0.04±0.01) |

**Table 1: Number of selective electrodes**

For the experiment and frequency bands reported in the main text, this table shows the number of electrodes selective to whole images ('Whole') or to both whole and partial images ('Both'). Also reported is the number of selective electrodes found when the object category labels were shuffled (mean±s.e.m., n=1000 iterations).

**Figure Legends**

**Figure 1: Experimental design and behavioral performance**

**(A)** After 500 ms fixation, an image containing a whole object or a partial object was presented for 150 ms. Subjects categorized objects into one of five categories (5-Alternative Forced Choice) following a choice screen. Presentation order was pseudo-randomized.

**(B)** Example images used in the task. Objects were either unaltered (Whole) or presented through Gaussian bubbles (Partial). For 12 subjects, the background was a gray screen (Main experiment), and for 6 subjects the background was phase-scrambled noise (Variant experiment). In this example, the object is seen through 5 bubbles (18% of object area shown). The number of bubbles was titrated for each subject to achieve 80% performance. Stimuli consisted of 25 different objects belonging to five categories.

**(C)** Above, percentage of the object visible (mean±SD) for each subject in the Main experiment (left) and the contrast-normalized Variant (right). Below, percentage of correct trials (performance) for Whole (black) and Partial (gray) objects. Average performance for Partial trials was 79±7 %, mean±SD (dashed line), well above chance (solid line).

**Figure 2: Example physiological responses from Main experiment**

Example responses from an electrode in the left Fusiform Gyrus.

**(A)** Intracranial field potential (IFP) responses to an individual exemplar object. For the Whole condition, the average response (green) and single trial traces (gray) are shown. For the Partial condition, example single trial responses (green, n=1) to different partial images of the same exemplar (top row) are shown. The response peak time is marked on the x-axis. The dashed line indicates the stimulus onset time and the black bar indicates stimulus presentation duration (150 ms).

**(B)** Raster of the neural responses for Whole (left, 52 trials) and Partial (right, 395 trials) objects for the category that elicited the strongest responses (human faces). Rows represent individual trials. Dashed lines separate responses to the 5 face exemplars. The color indicates the IFP at each time point (bin size = 2 ms, see scale on top).

**(C)** Average IFP response to Whole (left) and Partial (right) objects belonging to five different categories (animals, chairs, human faces, fruits, and vehicles, see color map on top). Shaded areas around each line indicate s.e.m. The gray rectangle denotes the image presentation time (150 ms). The total number of trials is indicated on the bottom right of each subplot.

**(D)** Selectivity was measured by computing the F-statistic at each time point for Whole (black) and Partial (gray) objects. Arrows indicate the first time point when the F-statistic was greater than the statistical threshold (black dashed line) for 25 consecutive milliseconds.

**(E)** Decoding performance (mean±SD) using a linear multi-class discriminant algorithm in classifying trials into one of five categories. Arrows indicate the first time when decoding performance reached the threshold for statistical significance (black dashed line). Chance is 20% (blue dashed line).

**(F)** Distribution of the visual response latency across trials for Whole (black) and Partial (gray) objects, based on when the IFP in individual trials was significantly above baseline activity. The distribution is based on kernel density estimate (bin size = 6 ms). The arrows denote the distribution means.

**Figure 3: Second example of physiological responses from Variant experiment**
Example responses from an electrode in the left Inferior Temporal Gyrus. The format and conventions are as in **Figure 2**, except that only three categories were tested, and the Partial Fixed condition was added in part **A** and **B** (**Experimental Procedures**). Note that the statistical thresholds for the F-statistic and decoding performance differ from those in **Figure 2** because of the different number of categories.

**Figure 4: Neural responses remained visually selective despite partial information**
**(A)** Average IFP amplitude $A = (1/N) \sum_{i=1}^{i=N} \max(IFP_i(t)) - \min(IFP_i(t))$ across trials ($N$) in response to partial versus whole objects for electrodes that were visually selective in the Whole condition (blue, n=61+22), and electrodes that were visually selective in both Whole and Partial conditions (gray, n=22) (Main experiment). Most of the data clustered around the diagonal (dashed line). Inset, distribution of suppression index: $(A_{whole} - A_{partial})/A_{whole}$.

**(B)** Locations of electrodes that showed visual selectivity in both Whole and Partial conditions. Example electrodes from **Figure 2** and **3** are marked by arrows. Colors indicate different brain gyri.

**(C)** Percent of total electrodes in each region that were selective in either the Whole condition (black) or in both conditions (gray). Color in the location name corresponds to the brain map in part **E**. The number of selective electrodes is shown next to each bar. Only regions with at least one electrode selective in both conditions are shown.

**(D)** For all pairs of discriminable trials (n = 10,438 pairs from 22 selective electrodes), we computed the distribution of the percent overlap in shared pixels. The percent overlap between two pairs of trials (inset, red and blue bubbles) was defined as the number of shared pixels (black) divided by the total object area (area inside gray outline).

**(E)** Latency of selective responses, as measured through ANOVA (e.g. **Figure 2D**) for electrodes selective in both Whole and Partial conditions from the Main (black, n=22) and Variant (gray, n=8) experiments. The latency distributions were significantly different (signed-rank test, main experiment: $p < 10^{-5}$, variant experiment: $p = 0.02$).

**(F)** Latency as measured by the machine-learning decoding analysis (e.g. **Figure 2E**). These latency distributions were significantly different (signed-rank test, main experiment: $p < 10^{-5}$, variant experiment: $p = 0.004$).

**Figure 5: Increased response latency for object completion**

**(A)** Distribution of visual response latencies in single trials for Whole (black) and Partial (gray) objects (as illustrated in **Figure 2F**). These distributions were significantly different (rank-sum test, $p<10^{-15}$). The vertical dashed lines denote the means of each distribution.

**(B)** Brain map of electrodes selective in both conditions, colored by the difference in the response latency (Partial – Whole; see color scale on the bottom).

**(C)** Comparison of response latency differences (Partial – Whole) between electrodes in occipital lobe (early visual) and temporal lobe (late visual).

**Figure 6: Summary of Latency Measurements**

**(A)** Decoding performance from pseudopopulation of 60 electrodes for categorization (thick lines) or exemplar identification (dotted lines) for Whole (black) or Partial (gray) conditions (**Experimental Procedures**). Horizontal lines indicate chance for categorization (20%) and identification (4%). Error bars represent standard deviation. The 60 electrodes used in this analysis were selected using their rank-order based on their individual decoding performance on training data.

**(B)** Summary of latency difference between Partial and Whole conditions for multiple definitions of latency (parentheses mark the figure source). Positive values means increased latency in the Partial condition. Box plots represent the median and quartile across the selective electrodes. For the Variant experiment, individual electrodes are plotted since the total number of electrodes *n* is small.

# Figure 1

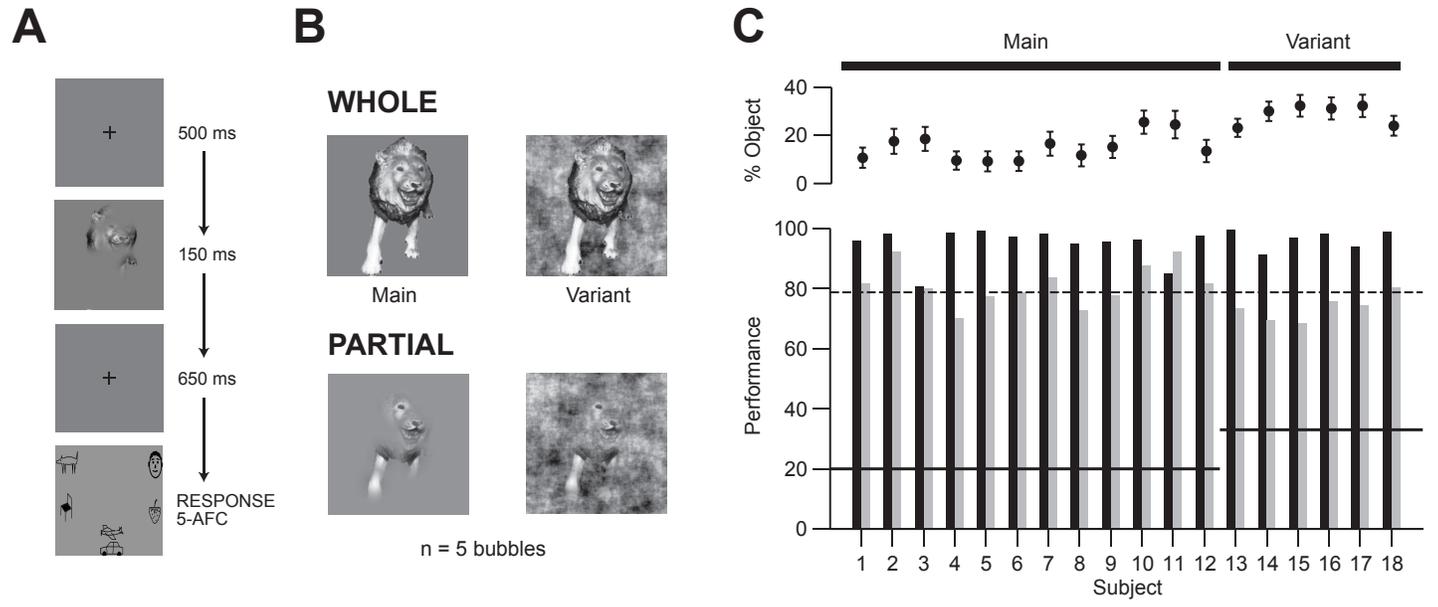

Figure 2

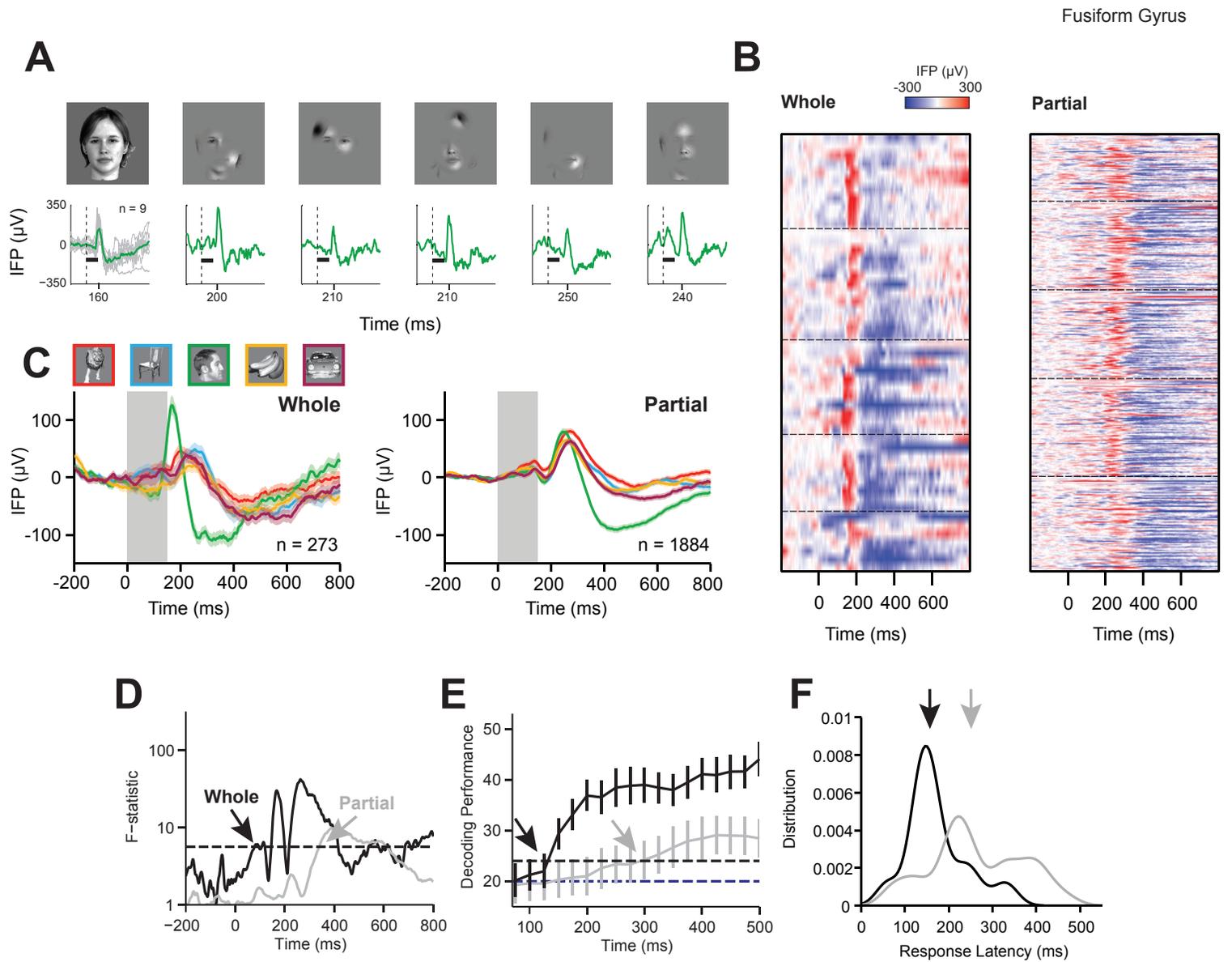

Figure 3

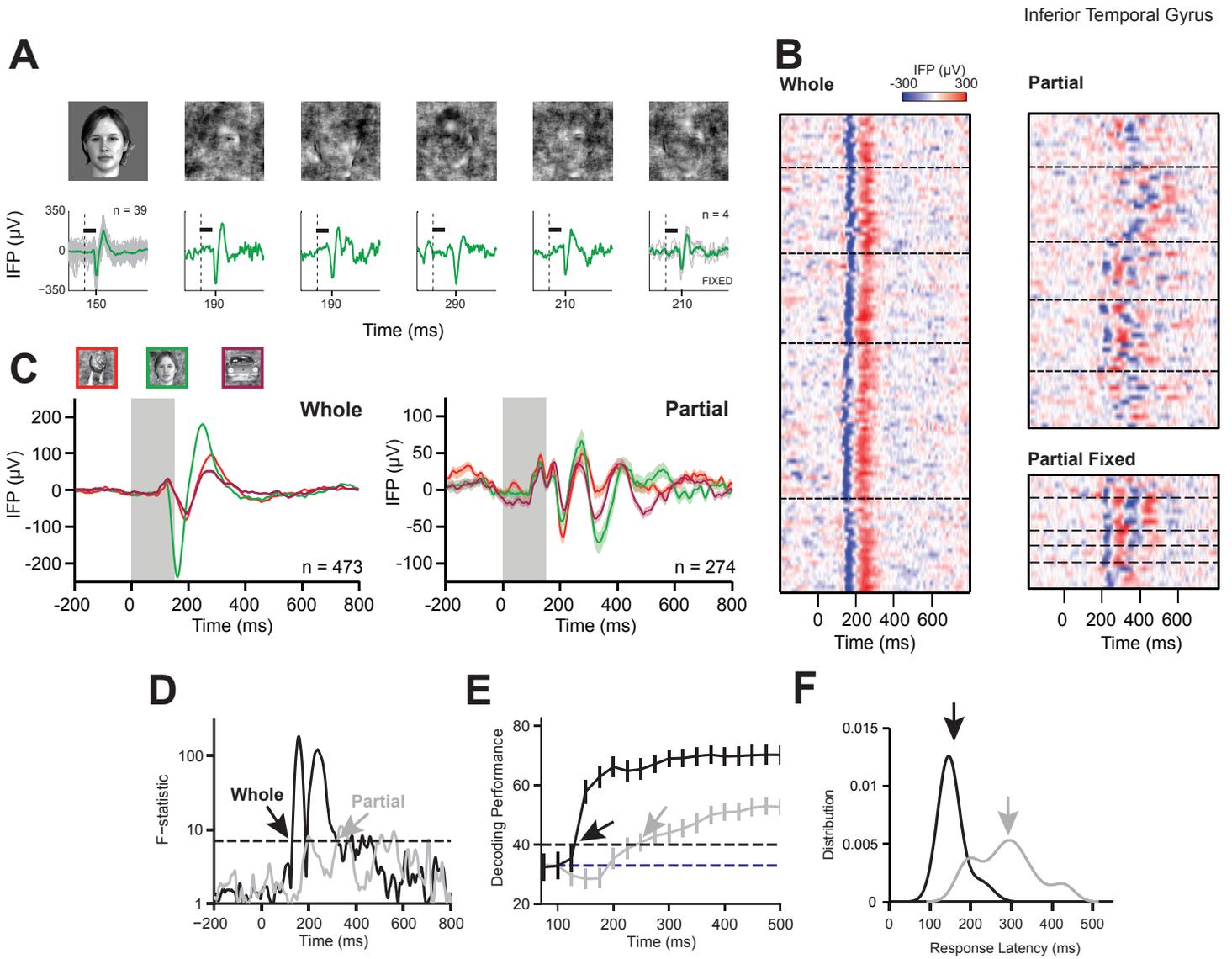

# Figure 4

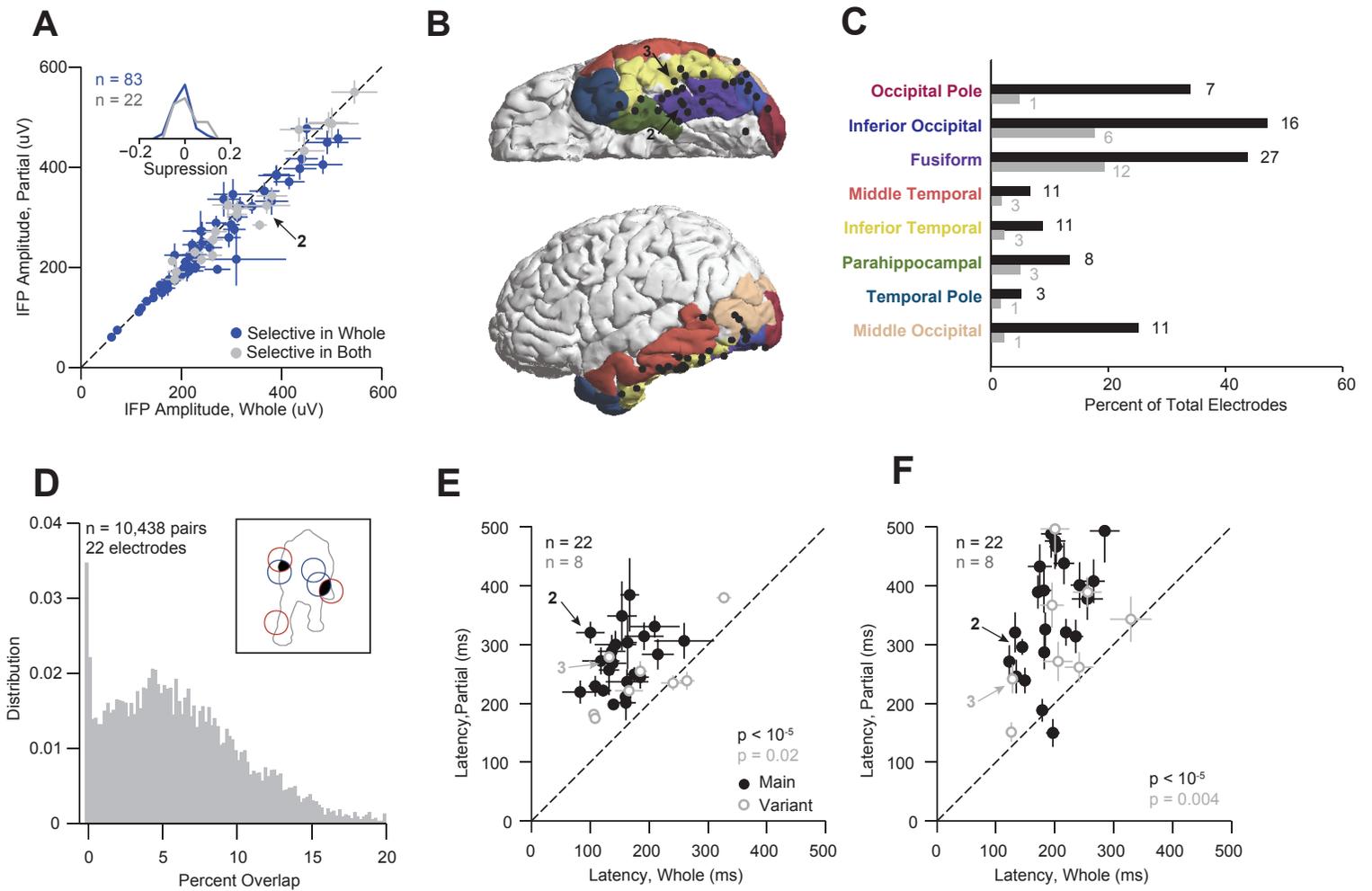

# Figure 5

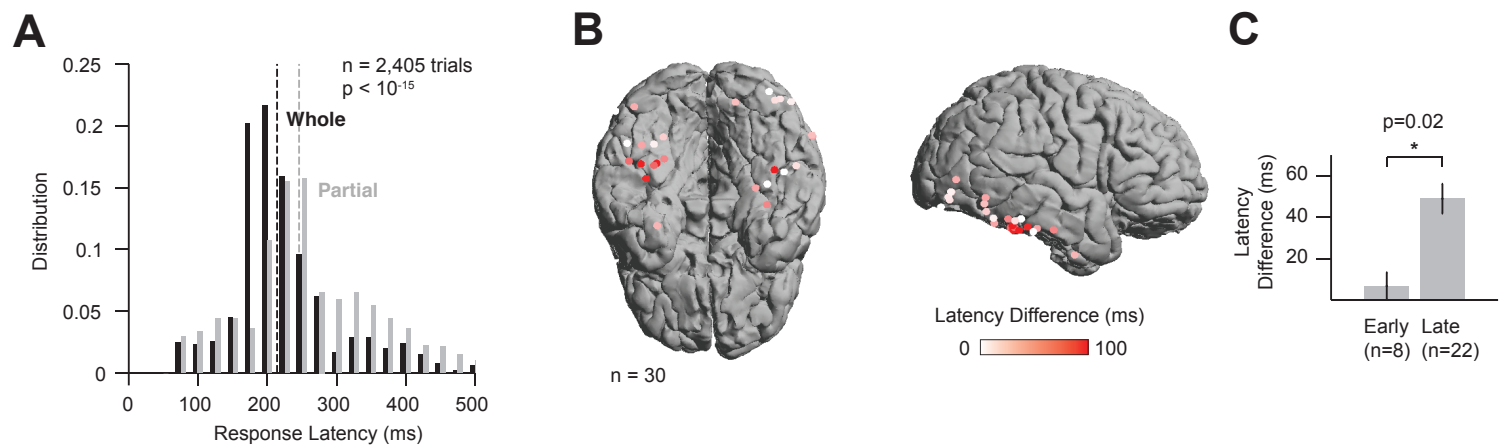

Figure 6

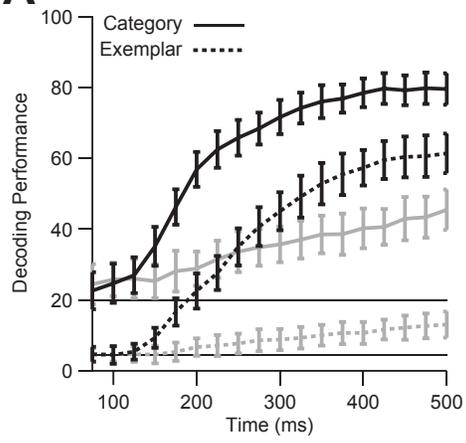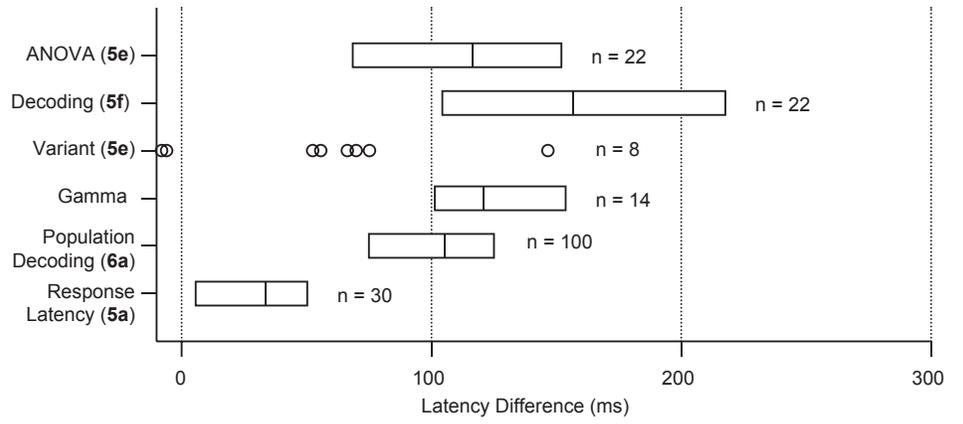

# Figure 7

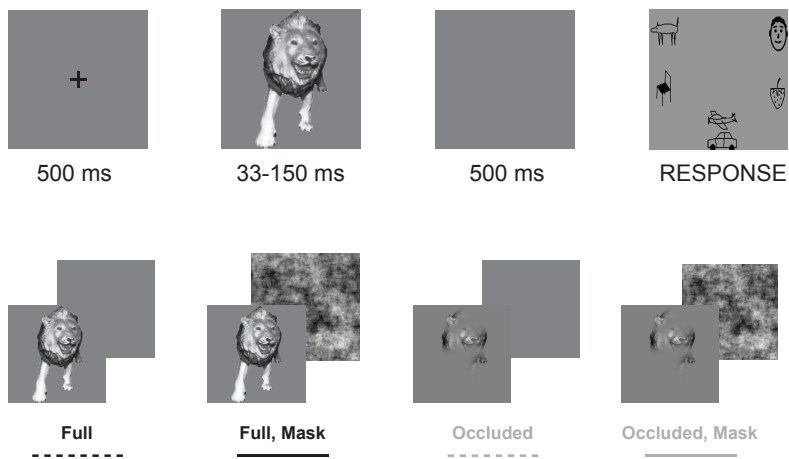
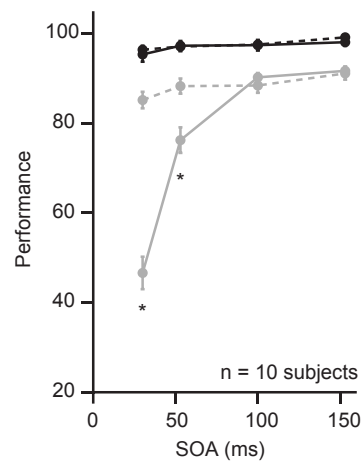
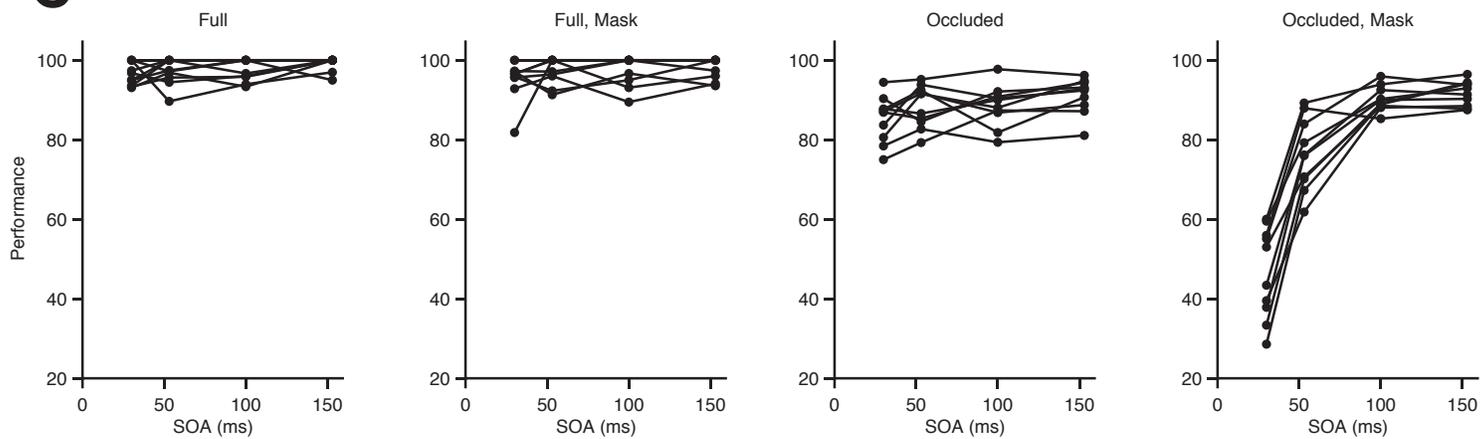

# Figure 8

**A**

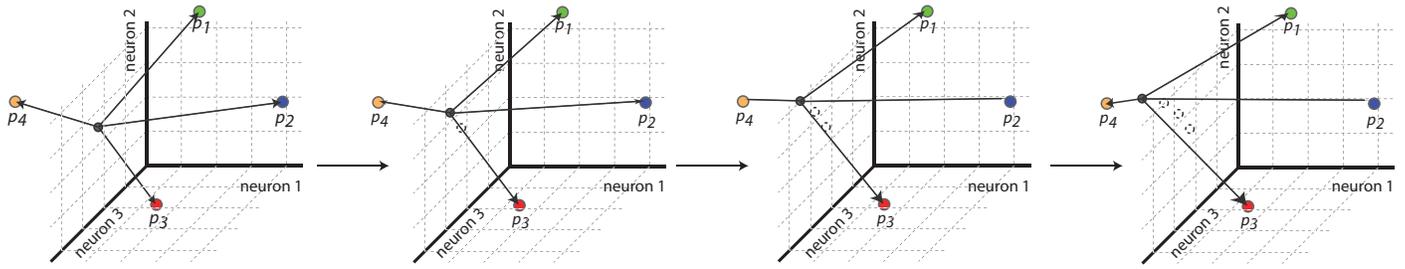

**B**

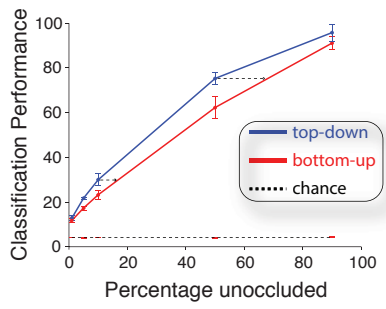